\documentclass[preprint]{JHEP3}

\usepackage{amsmath}
\usepackage{epsfig}
\newcommand{\ep}{\varepsilon}

\newcommand{\Li}[2]{{\mbox{Li}}_{#1}\left(#2\right)}
\newcommand{\Cl}[2]{{\mbox{Cl}}_{#1}\left(#2\right)}
\newcommand{\Ls}[2]{{\mbox{Ls}}_{#1}\left(#2\right)}
\newcommand{\LS}[3]{{\mbox{Ls}}_{#1}^{(#2)}\left(#3\right)}
\newcommand{\Lsc}[2]{{\mbox{Lsc}}_{#1\!}\left(#2\right)}

\newcommand{\Snp}[2]{{\mbox{S}}_{#1\!}\left(#2\right)}
\newcommand{\Ti}[2]{{\mbox{Ti}}_{#1}\left(#2\right)}

\title{All order $\ep$-expansion of Gauss hypergeometric functions with 
integer and half/integer values of parameters
~\footnote{Supported by 
NATO Grant PST.CLG.980342
and DOE grant DE-FG02-05ER41399}
      }
\author{
M.~Yu.~Kalmykov \\
Department of Physics, Baylor University, \\
One Bear Place, Box 97316, Waco, TX 76798-7316 \\
\\
Bogoliubov Laboratory of Theoretical Physics, Joint Institute for Nuclear Research, \\
$141980$ Dubna (Moscow Region), Russia \\
Email: \email{kalmykov@theor.jinr.ru}
}
\author{
B.F.L.~Ward, \quad S.~Yost, \\ 
Department of Physics, Baylor University, \\
One Bear Place, Box 97316 Waco, TX 76798-7316}

\keywords{
Gauss hypergeometric functions,
harmonic polylogarithms, 
colour polylogarithms, 
Laurent expansion of Gauss hypergeometric function, multiloop calculations}

\preprint{ \\ hep-th/0612240 \\ BU-HEPP-06-12}

\date{December 2006}

\abstract{ \\
It is proved that the Laurent expansion of 
the following Gauss hypergeometric functions, 
\begin{eqnarray}
&{}_2F_{1}&(I_1+a\ep, I_2+b\ep; I_3+c \ep;z) \;,
\nonumber \\ 
&{}_2F_{1}&(I_1+a\ep, I_2+b\ep; I_3+\tfrac{1}{2}+c \ep;z) \;,
\nonumber \\ 
&{}_2F_{1}&(I_1+\tfrac{1}{2}+a\ep,   I_2+b\ep; I_3+c \ep;z) \;,
\nonumber \\ 
&{}_2F_{1}&(I_1+\tfrac{1}{2}+a\ep,   I_2+b\ep; I_3+\tfrac{1}{2} + c \ep;z) \;,
\nonumber \\ 
&{}_2F_{1}&(I_1+\tfrac{1}{2}+a\ep, I_2+\tfrac{1}{2}+b\ep; I_3+\tfrac{1}{2} + c \ep;z) \;,
\nonumber 
\end{eqnarray}
where $I_1,I_2,I_3$ are an arbitrary integer nonnegative numbers,
$a,b,c$ are an arbitrary numbers 
and $\ep$ is an arbitrary small parameters, 
are expressible in terms of the harmonic 
polylogarithms of Remiddi and Vermaseren with polynomial coefficients.
An efficient algorithm for the calculation of 
the higher-order coefficients of Laurent expansion is constructed.
Some particular cases of Gauss hypergeometric functions are also discussed. 
}

\begin{document}

\tableofcontents

\section{Introduction}

One of the most powerful techniques for calculating  Feynman diagrams 
is based on 
their presentation in terms of hypergeometric functions. 
We will call this the hypergeometric function representation 
of Feynman diagrams.
Such a representation can be used for numerical evaluation, 
construction of the asymptotic expansion, {\it etc}. 
One of the unsolved problems in this program 
is obtaining the proper representation for a diagram with an 
arbitrary number of legs and loops. 
Direct use of $\alpha$- or Feynman parameters representations \cite{Bogolyubov}
is not very helpful in solving this problem. 
The Mellin-Barnes technique is restricted to several 
topologies \cite{BD,DK01,JKV02,JK04}. 
The negative dimension approach \cite{nda} has a similar 
restriction \cite{AGO:nda}.
The most investigated diagrams are 
the master integrals (typically, integrals with the power of each propagator 
equal to unity). 
The differential \cite{DE} and/or difference equation \cite{Tarasov00} 
techniques are usually used to obtain such representations. 
The known cases include the one-loop diagrams \cite{one-loop,davydychev}, 
two-loop propagator-type diagrams with special 
mass and momentum values \cite{BFT}, 
several three-loop bubble-type diagrams \cite{DK01}, 
three-loop vertex-type diagrams \cite{3-vertex}, 
and four-loop bubble-type diagrams \cite{4-loop}.
For practical application however, it is necessary to construct the 
of $\ep$-expansion (Laurent expansion) of hypergeometric functions. 
There is some evidence that the multiple polylogarithms 
\cite{Goncharov,Broadhurst:1998,Borwein:1999}, 
\begin{equation}
\Li{k_1,k_2, \cdots, k_n}{z_1,z_2,\cdots, z_n} = 
\sum_{m_1 > m_2 > \cdots m_n > 0} \frac{z_1^{m_1} z_2^{m_2} \cdots z_n^{m_n} }{m_1^{k_1} m_2^{k_2} \cdots m_n^{k_n}} \;.
\end{equation}
are sufficient for parametrizing the coefficients of the $\ep$-expansion 
of some, but not all\footnote{We are thankful to S. Weinzierl for this information.},  
hypergeometric functions \cite{nested2}.

In some particular cases, the result of the Laurent expansion can be written 
in terms of simpler functions. 
In particular, at the present moment, it is commonly accepted 
\cite{weinzierl:03} that the generalized hypergeometric functions with 
an arbitrary set of integer parameters can be 
presented in terms of harmonic polylogarithms \cite{RV00}. 
The idea of the proof is based on the 
properties of {\it nested sums} \cite{nested1}: the analytical 
coefficients of the $\ep$-expansion of any generalized hypergeometric 
function with integer parameters can be reduced to a set of some 
basic {\it harmonic} series of the type 
\begin{eqnarray}
\hspace{-5mm}
\sum_{j=1}^\infty \frac{z^j}{j^c} S_{a_1}(j-1) \cdots S_{a_p}(j-1) \;, 
\label{harmonic}
\end{eqnarray}
where $z$ is an arbitrary argument and 
$S_a(j)$ is an harmonic sum defined as $S_a(j) = \sum_{k=1}^j \frac{1}{k^a}$.
Series of this type are expressible in terms of the 
Remiddi-Vermaseren harmonic polylogarithms.

However for hypergeometric functions with half-integer values of parameters, 
the new type of sums, {\it multiple} ({\it inverse}) {\it binomial sums} 
\cite{KV00,JKV02,JK04,DK04} are generated:
\begin{eqnarray}
\hspace{-5mm}
\Sigma^{(k)}_{a_1,\cdots,a_p; \; b_1,\cdots,b_q;c}(z)
&\equiv&
\sum_{j=1}^\infty \frac{1}{\left(2j \atop j\right)^k}\frac{z^j}{j^c}
S_{a_1}(j-1) \cdots S_{a_p}(j-1) S_{b_1}(2j-1) \cdots S_{b_q}(2j-1) \; .
\nonumber\\
& &
\label{binsum}
\end{eqnarray}
For particular values of $k$, the sums (\ref{binsum}) are called 
\begin{eqnarray}
k = 
\left\{ 
\begin{array}{rl}
 0  & \mbox{ {\it generalized harmonic} } \\
 1  & \mbox{ {\it inverse binomial} } \\
-1  & \mbox{ {\it binomial} } 
\end{array} \right\} \mbox{ sums }.
\nonumber 
\end{eqnarray}

At the present moment, there is no proof 
that any {\it multiple} ({\it inverse}) {\it binomial sums}
can be expressed in terms of harmonic polylogarithms only. 
This problem was investigated in Ref.\ \cite{DK04} for {\it multiple inverse binomial sums} up to {\it weight 4}. 
In Ref.\ \cite{MKL04}, it was shown that some of the {\it multiple inverse binomial sums}
are not expressible in terms of harmonic polylogarithms of simple argument. 
In Ref.\ \cite{MKL06}, the results of Ref.\ \cite{DK04} were extended on the case of special combinations of 
{\it multiple binomial sums} and  {\it multiple generalized harmonic sums}.
However, the Laurent expansion of a hypergeometric function in general contains
combinations of {\it multiple sums}. These combinations may be expressed in 
terms of harmonic polylogarithms. 
From this point of view, the construction of the analytical coefficients of 
the $\ep$-expansion of hypergeometric functions can be done independently 
from existing results for each individual {\it multiple sum}.\footnote{We are indebted to A.~Davydychev for 
discussion on this subject.}

The simplest hypergeometric function is the
Gauss hypergeometric function $_{2}F_1(a,b;c;z)$, \cite{Gauss,bateman,Slater}.
It satisfies the second-order differential equation 
\begin{eqnarray}
\frac{d}{dz} 
\left( 
z \frac{d}{dz} + c - 1
\right)w(z)
= 
\left( z \frac{d}{dz} + a \right)
\left( z \frac{d}{dz} + b  \right)w(z) \;,
\end{eqnarray}
and admits the series representation
\begin{eqnarray}
{}_{2}F_1(a,b;c;z)
= \sum_{k=0}^\infty \frac{(a)_k (b)_k}{(c)_k} \frac{z^k}{k!} \;, 
\end{eqnarray}
where $(a)_k = \Gamma(a+k)/\Gamma(a)$ is the Pochhammer symbol.

The primary aim of this paper is to prove the following:

\noindent 
\begin{itemize}
\item
{\bf Theorem 1:} \\
\ {\it 
The all-order $\ep$-expansions of the Gauss hypergeometric functions 
\begin{subequations}
\label{2F1-Theorem1}
\begin{eqnarray}
&{}_2F_{1}&(I_1+a\ep, I_2+b\ep; I_3+c \ep;z) \;,
\\ 
&{}_2F_{1}&(I_1+a\ep, I_2+b\ep; I_3+\tfrac{1}{2}+c \ep;z) \;,
\\ 
&{}_2F_{1}&(I_1+\tfrac{1}{2}+a\ep,   I_2+b\ep; I_3+c \ep;z) \;,
\\ 
&{}_2F_{1}&(I_1+\tfrac{1}{2}+a\ep,   I_2+b\ep; I_3+\tfrac{1}{2} + c \ep;z) \;,
\\ 
&{}_2F_{1}&(I_1+\tfrac{1}{2}+a\ep, I_2+\tfrac{1}{2}+b\ep; I_3 + \tfrac{1}{2} + c \ep;z) \;,
\end{eqnarray}
\end{subequations}
where $\{ I_k \}$ are integer numbers, $a,b,c$ are an arbitrary numbers,  
and $\ep$ is an arbitrary small parameter, 
are expressible in terms of Remiddi-Vermaseren harmonic 
polylogarithms with rational coefficients. }
\end{itemize}

\section{All-order $\ep$-expansion}
\label{allorder}
\subsection{Non-zero values of the $\ep$-dependent part}

It is well known that any Gauss hypergeometric function 
may be expressed as a linear combination of two other hypergeometric 
functions with parameters differing from the 
original ones by an integer \cite{Gauss,bateman,Slater,nikiforov1,MKL06}. 
Such a representation will be called a {\it reduction}, and the explicit 
algorithm will be called a {\it reduction algorithm}. 
Using the algorithm described in Ref.\ \cite{MKL06}, the result of the 
reduction can be written as 
\begin{eqnarray}
&& \hspace{-5mm}
P(a,b,c,z)
{}_{2}F_{1}(a+I_1,b+I_2;c+I_3;z)
 = 
\Biggl \{
  Q_1(a,b,c,z) \frac{d}{dz}
+ Q_2(a,b,c,z) 
\Biggr\}
{}_{2}F_{1}(a,b;c; z) \;,
\nonumber \\ 
\label{decomposition}
\end{eqnarray}
where 
$a,b,c,$ are any fixed numbers,  
$P,Q_1,Q_2$ are polynomial in parameters $a,b,c$ and argument $z$, 
and $I_1,I_2,I_3$ any integer numbers. 

All the hypergeometric functions (\ref{2F1-Theorem1}) listed in {\bf Theorem 1}
can be reduced to functions with $I_1,I_2,I_3$ equal to zero for half-integer 
values of parameters, and to unity for integer ones.
In this way, all the hypergeometric functions of {\bf Theorem 1} 
are expressible in terms of the
following five basic functions and their first derivatives: 
\begin{subequations}
\label{2F1}
\begin{equation}
\label{2F1-type1}
{}_2F_{1}(a_1\ep, a_2\ep; 1+c \ep;z), \qquad 
{}_2F_{1}(a_1\ep, a_2\ep; \tfrac{1}{2}+f \ep;z), 
\end{equation}
\begin{equation}
\label{2F1-type2}
{}_2F_{1}(\tfrac{1}{2}\!+\!b\ep,   a\ep; 1\!+\!c \ep;z), \quad 
{}_2F_{1}(\tfrac{1}{2}\!+\!b\ep,   a\ep; \tfrac{1}{2} \!+\! f \ep;z), \quad 
{}_2F_{1}(\tfrac{1}{2}\!+\!b_1\ep, \tfrac{1}{2}\!+\!b_2\ep; \tfrac{1}{2} \!+\! f \ep;z). 
\end{equation}
\end{subequations}
It was shown in Ref.\ \cite{MKL06} that only the two hypergeometric functions 
(and their first derivative) of type (\ref{2F1-type1}) 
are algebraically independent.  The other three, (\ref{2F1-type2}),
are algebraically expressible in terms of 
${}_2F_{1}(a_1\ep, a_2\ep; \tfrac{1}{2}+f \ep;z)$. 
Consequently, in order to prove {\bf Theorem 1}, it sufficient to show that
the analytical coefficients of the $\ep$-expansion of the first two 
hypergeometric functions (\ref{2F1-type1}) are
expressible in terms of Remiddi-Vermaseren polylogarithms. 
\subsubsection{Integer values of of $\ep$-independent parameters}
Let us start from expansion of Gauss hypergeometric functions with integer values of parameters, 
and consider the function
${}_2F_{1}(a_1\ep, a_2\ep; 1+c \ep;z)$.
In ref.\ \cite{hyper:expansion}, the all-order $\ep$-expansions for this 
functions and its first derivative were constructed
in terms of multiple polylogarithms of one variable \cite{Broadhurst:1998,Borwein:1999}. 
These multiple polylogarithms may be expressed as iterated integrals
\footnote{
Recall that multiple polylogarithms can be expressed as iterated integrals of
the form
\begin{eqnarray}
\Li{k_1, \cdots, k_n}{z} & = &  
\int_0^z 
\underbrace{\frac{dt}{t} \circ \frac{dt}{t} \circ \cdots \circ \frac{dt}{t}}_{k_1-1 \mbox{ times}} \circ \frac{dt}{1-t} 
\circ \cdots \circ 
\underbrace{\frac{dt}{t} \circ \frac{dt}{t} \circ \cdots \circ \frac{dt}{t}}_{k_n-1 \mbox{ times}} \circ \frac{dt}{1-t}
\;,  
\label{iterated}
\end{eqnarray}
where, by definition
\begin{eqnarray}
\int_0^z 
\underbrace{\frac{dt}{t} \circ \frac{dt}{t} \circ \cdots \circ \frac{dt}{t}}_{k_1-1 \mbox{ times}} \circ \frac{dt}{1-t} 
= 
\int_0^z 
\frac{dt_1}{t_1} \int_0^{t_1} \frac{dt_2}{t_2} \cdots \int_0^{t_{k-2}} \frac{dt_{k_1-1}}{t_{k_1-1}}
\int_0^{t_{k_1-1}} \frac{dt_{k_1}}{1-t_{k_1}} \;.
\end{eqnarray}
The integral (\ref{iterated}) is an iterated Chen integral \cite{Chen} (see also \cite{dirk})
w.r.t.\ the two differential forms $\omega_0 = dz/z$ and $\omega_1 = \frac{dz}{1-z}$, so that 
\begin{eqnarray}
\Li{k_1, \cdots, k_n}{z} & = & \int_0^z \omega_0^{k_1-1} \omega_1 \cdots \omega_0^{k_n-1} \omega_1 \;.
\label{chen}
\end{eqnarray}
} 
and have the expansion
\begin{equation}
\Li{k_1,k_2, \cdots, k_n}{z} = 
\sum_{m_1 > m_2 > \cdots m_n > 0} \frac{z^{m_1}}{m_1^{k_1} m_2^{k_2} \cdots m_n^{k_n}} \;.
\label{mp}
\end{equation}
Similar results were also derived (without explicit form of coefficients) 
in Ref.\ \cite{weinzierl:03} via nested sums approach. 
We will follow the idea of Ref.\ \cite{hyper:expansion}.

The Gauss hypergeometric function ${}_2F_{1}(a_1\ep, a_2\ep; 1+c \ep;z)$
is the solution of the differential equation 
\begin{eqnarray}
\frac{d}{dz} 
\left( z \frac{d}{dz} + c \ep   \right) w(z)
= 
\left( z \frac{d}{dz} + a_1 \ep \right) \left( z \frac{d}{dz} + a_2 \ep\right) w(z) \;,
\label{gauss:diff}
\end{eqnarray}
with boundary conditions $w(0)=1$ and $\left. z \frac{d}{dz} w(z)\right|_{z=0} = 0$.
Eq.\ (\ref{gauss:diff}) is valid in each order of $\ep$, so that 
in terms of coefficients functions $w_k(z)$ defined as 
\begin{equation}
w(z) = \sum_{k=0}^\infty w_k(z) \ep^k, 
\label{epsilon-expansion}
\end{equation}
it can be written 
\begin{eqnarray}
(1-z) \frac{d}{dz} 
\left( z \frac{d}{dz} \right) w_k(z)
= 
\left( a_1 + a_2 - \frac{c}{z} \right) \left( z \frac{d}{dz} \right) w_{k-1}(z) 
+ a_1 a_2 w_{k-2}(z)
\label{gauss:diff2}
\end{eqnarray}
for $k \geq 0$ with  
\begin{subequations}
\label{wspecial}
\begin{eqnarray}
\label{w0}
 & w_0(z) &= 1\;,  \\
 & w_k(z) &= 0\;, \qquad k<0.
\label{wneg}
\end{eqnarray}
\end{subequations}
The boundary conditions for the coefficient functions are
\begin{subequations}
\label{boundary}
\begin{eqnarray}
\label{boundary1}
 & w_k(0) = 0\;,& \qquad  k \geq 1 \;, \\
& \left. z \frac{d}{dz} w_k(z) \right|_{z=0} = 0\;,& \qquad  k \geq 0 \; .
\label{boundary2}
\end{eqnarray}
\end{subequations}
Let us introduce a new function $\rho(z)$ defined by\footnote{
We may note that 
$$
{}_{2}F_1\left(\begin{array}{c|}
1+a_1\ep, 1\!+\!a_2\ep\\
2  \!+\! c \ep  \end{array} ~z \right) 
= 
\frac{1+c\ep}{z}
\sum_{k=0}^\infty 
\left[ \frac{\rho_{k+2}(z)}{a_1 a_2} \right]
\ep^k \;.
$$
} 
\begin{equation}
\rho(z)  =  z \frac{d}{dz} w(z) = \sum_{k=0}^\infty \rho_k(z) \ep^k \;,
\end{equation}
where the coefficient functions satisfy 
\begin{equation}
\rho_k(z) = z \frac{d}{dz} w_k(z) \;.
\end{equation}
The boundary conditions for these new functions follow from 
Eq.\ (\ref{boundary}):
\begin{equation}
\rho_k(0)  =  0 \;, \qquad k \geq 0 \;.
\label{boundary:rho}
\end{equation}
Eq.\ (\ref{gauss:diff2}) can be rewritten as a system of two first-order 
differential equations:
\begin{eqnarray}
(1-z) \frac{d}{dz} \rho_i (z)
& = &  
\left(a_1 \!+\! a_2 \!-\! \frac{c}{z} \right) \rho_{i-1}(z)
\!+\! a_1 a_2 w_{i-2}(z) \;,
\nonumber \\ 
z \frac{d}{dz} w_i(z) & = & \rho_i(z) \;.
\label{gauss:diff3}
\end{eqnarray}
The solution of this system can be presented in an iterated form: 
\begin{eqnarray}
\rho_i (z)
& = &  
\left(a_1 \!+\! a_2 \!-\! c \right) \int_0^z \frac{dt}{1-t} \rho_{i-1}(t)
\!+\! a_1 a_2 \int_0^z \frac{dt}{1-t} w_{i-2}(t) 
\!-\! c w_{i-1}(z) \;, \quad i \geq 1 \;,
\nonumber \\ 
w_i (z) & = & \int_0^z \frac{dt}{t} \rho_i(t) \;, \quad i \geq 1 \;.
\label{w}
\end{eqnarray}
Taking into account that $w_0(z)=1$ and $\rho_0(z)=0$ (the $\ep$-expansion of $\rho(z)$ 
begins with the term linear in $\ep$), we obtain the first few coefficients,
\begin{subequations}
\label{first}
\begin{eqnarray}
\rho_1(z) &=& w_1(z) = 0, \\
\frac{\rho_2(z)}{a_1 a_2} &=& - \ln(1-z) \equiv H(1;z) \;,\\
\frac{w_2(z)}{a_1 a_2} &=& \Li{2}{z} \equiv H(0,1;z) \;,\\
\frac{\rho_3(z)}{a_1 a_2} &=&
\gamma_c \frac{1}{2} \ln^2(1-z) - c \Li{2}{z} 
\equiv \gamma_c  H(1,1;z) \!-\! c H(0,1;z)\;,\\
\frac{w_3(z)}{a_1 a_2 } &=&
\gamma_c \Snp{1,2}{z} \!-\! c \Li{3}{z} 
\equiv \gamma_c  H(0,1,1;z) \!-\! c H(0,0,1;z)\;,
\end{eqnarray}
\end{subequations}
where we have defined $\gamma_c = a_1 + a_2 - c$,
and 
$\Li{n}{z}$ and $S_{a,b}(z)$ are the classical and Nielsen polylogarithms 
\cite{Lewin,Nielsen}, respectively: 
$$
S_{a,b}(z)  =  \frac{(-1)^{a+b-1}}{(a-1)! \; b!} \!\int\limits_0^1
\mbox{d} \xi\; \frac{\ln^{a-1}\!\xi \ln^b (1\!-\!z\xi)}{\xi} \; ,
\quad 
S_{a,1}(z)  =  \Li{a+1}{z}.
$$
The functions $H(\vec{A};z)$ are the 
Remiddi-Vermaseren harmonic polylogarithms \cite{RV00}, 
and $\vec{A}$ is a multiple index including only entries $0$ and $1$.

From the representation (\ref{w}) and result for the first few coefficients 
(\ref{first}) we may derive the following observations: 
\begin{itemize}
\item
{\bf Corollary 1:}
{\it 
The all-order $\ep$-expansion of the function 
${}_2F_{1}(a_1 \ep, a_2 \ep; 1+c \ep;z)$
may be written in terms of 
harmonic polylogarithms $H_{\vec{A}}(z)$ only, where the multiple index $\vec{A}$ includes only the values $0$ and $1$. 
}

\item
{\bf Corollary 2:}
{\it 
The analytical coefficient of $\ep^k$ in the expansion of
${}_2F_{1}(a_1 \ep, a_2 \ep; 1+c \ep;z)$
includes only 
functions of weight k with numerical coefficients. 
}

\item
{\bf Corollary 3:}
{\it 
The non-constant terms of the $\ep$-expansion of 
${}_2F_{1}(a_1 \ep, a_2 \ep; 1+c \ep;z)$
are proportional to the product $a_1 a_2$ in any order of $\ep$.
}
\end{itemize}
The first and the last statement follows from the representation\footnote{The {\bf Corollary 3} follows also 
from general properties of hypergeometric functions.} (\ref{w}), 
the explicit value of coefficients functions $w_k(z)$, $k=0,1,2$ (Eqs.\ (\ref{first})), 
and definition of harmonic polylogarithms \cite{RV00}.
The second statement follows from the form of the solution of Eq.~(\ref{w}).

The relation between harmonic polylogarithms $H_{\vec{A}}(z)$, with multiple index $\vec{A}$ including only $0$ and $1$, 
and multiple polylogarithms of one variable (Eq.\ \ref{mp}))
is well known \cite{nested1} and follows from the proper definition 
(see Sec.\ 2 in Ref.\ \cite{RV00}):
\begin{eqnarray}
\Li{k_1, k_2, \cdots, k_n}{z} & = &  
H( \underbrace{0,0, \cdots, 0,}_{k_1-1 \mbox{ times}} 1,
   \underbrace{0,0, \cdots, 0,}_{k_2-1 \mbox{ times}} 1,
   \cdots 
   \underbrace{0,0, \cdots, 0,}_{k_n-1 \mbox{ times}} 1;z) \;.
\end{eqnarray}

By continued iterations of Eq.\ (\ref{w}) and Eq.\ (\ref{first}), 
we have reproduced all coefficients of the $\ep$-expansion of the 
Gauss hypergeometric function 
presented in Eq.\ (4.7) of \cite{MKL06}. For the coefficients functions 
$\rho_4(z), \omega_4(z)$, and $\rho_5(z)$, we find a more compact form.
We also obtain the higher-order terms $\omega_5(z)$ and $\rho_6(z)$
of the $\ep$-expansion. The results are\footnote{The FORM\cite{FORM} 
representation of these expressions can be extracted from Ref.\ \cite{MKL}.}
\begin{eqnarray}
&& 
\frac{\rho_4(z)}{a_1 a_2} = 
- \frac{1}{6} \gamma_c^2 \ln^3(1 \!-\! z)
\!+\! \left( c \gamma_c \!-\! a_1 a_2 \right) \ln(1 \!-\! z) \Li{2}{z}
\!+\! c^2 \Li{3}{z}
\!+\! \left( c \gamma_c \!-\! 2 a_1 a_2 \right) \Snp{1,2}{z} \;,
\nonumber \\ && 
\frac{w_4(z)}{a_1 a_2 } = 
c^2 \Li{4}{z}
- \frac{1}{2} \left( c \gamma_c - a_1 a_2 \right)\left[ \Li{2}{z} \right]^2
+ \gamma_c^2 \Snp{1,3}{z}
+ \left( c \gamma_c - 2 a_1 a_2 \right) \Snp{2,2}{z} \;,
\label{first:new}
\\ && 
\frac{\rho_5(z)}{a_1 a_2} = 
\frac{1}{24} \gamma_c^3 \ln^4(1-z)
\!-\! c^3 \Li{4}{z}
\!-\! c \gamma_c^2 \Snp{1,3}{z} 
\nonumber \\ && \hspace{15mm}
- c \left( c \gamma_c \!-\! a_1 a_2 \right) \ln (1-z) \Li{3}{z} 
\!-\! c \left( c \gamma_c \!-\! 2 a_1 a_2 \right) \Snp{2,2}{z} 
\nonumber \\ && \hspace{15mm}
- \gamma_c \left( c \gamma_c \!-\! a_1 a_2 \right) \ln(1-z) 
\Biggl[ \frac{1}{2} \ln(1-z) \Li{2}{z} + \Snp{1,2}{z} \Biggr] \;,
\\ && 
\frac{w_5(z)}{a_1 a_2 } = 
\gamma_c^3 \Snp{1,4}{z}
- c^3 \Li{5}{z}
- c \gamma_c^2 \Snp{2,3}{z}
- c \left( c \gamma_c - 2 a_1 a_2 \right) \Snp{3,2}{z}
\nonumber \\ && \hspace{15mm}
+ \left( c \gamma_c - a_1 a_2 \right) 
\Biggl[
\gamma_c \Li{2}{z} \Snp{1,2}{z} - \gamma_c  F_1(z) - c F_2(z) 
\Biggr]
\;,
\\ && 
\frac{\rho_6(z)}{a_1 a_2} = 
- \gamma_c^4 \frac{1}{120} \ln^5 (1-z)
+ c^4 \Li{5}{z}
+ c^2 \gamma_c^2 \Snp{2,3}{z}
\nonumber \\ && \hspace{15mm}
+ \frac{1}{6} \gamma_c^2 \left( c \gamma_c - a_1 a_2 \right) 
\Biggl[
  \ln^3(1-z) \Li{2}{z}
+ 3 \ln^2(1-z) \Snp{1,2}{z}
+ 6 \ln(1-z) \Snp{1,3}{z}
\Biggr]
\nonumber \\ && \hspace{15mm}
+ \frac{1}{2} c \left( c \gamma_c - a_1 a_2 \right) \ln(1-z)
\Biggl[ \gamma_c \ln(1-z) \Li{3}{z} + 2 c \Li{4}{z} \Biggr]
\nonumber \\ && \hspace{15mm}
- (c-a_1) (c-a_2) \left( c \gamma_c - 2 a_1 a_2 \right)  \ln(1-z) \Snp{2,2}{z}
\nonumber \\ && \hspace{15mm}
+ a_1 a_2 \left( c \gamma_c - a_1 a_2 \right) 
\Biggl[
\frac{1}{2} \ln(1-z) \left[ \Li{2}{z} \right]^2
- 2 \Li{2}{z} \Snp{1,2}{z}
+ 2 F_1(z)
\Biggr]
\nonumber \\ && \hspace{15mm}
+ \left( c \gamma_c - 2 a_1 a_2 \right) 
\Biggl[ \gamma_c^2 \Snp{1,4}{z} + c^2 \Snp{3,2}{z} \Biggr] \;,
\label{first:new:last}
\end{eqnarray}
where we have introduced two new functions:
\begin{eqnarray}
F_1(z) & = & \int_0^z \frac{dx}{x} \ln^2(1 - x) \Li{2}{x} \; , 
\\
F_2(z) & = & \int_0^z \frac{dx}{x} \ln(1 - x) \Li{3}{x} \; .
\end{eqnarray}
There is an algebraic relation\footnote{We are indebted to A.~Davydychev for this relation.} 
between these two functions: 
\begin{eqnarray}
F_2(1-z) & = & 
  F_1(z)
- 2 \ln z  \Snp{1,3}{z} 
+ 2 \Snp{2,3}{z} 
- \Li{2}{z} \Snp{1,2}{z}
- \ln z \ln(1-z)  \Snp{1,2}{z} 
\nonumber \\ && 
- \frac{1}{6} \ln^3(1-z) \ln^2 z 
- \frac{1}{2} \ln z \ln^2 (1-z) \Li{2}{z}
+ \frac{1}{2} \zeta_2 \ln^2 (1-z) \ln z 
\nonumber \\ && 
- \zeta_2 \Snp{1,2}{z}
- \zeta_3 \Li{2}{1-z}
- \zeta_5 \;,
\end{eqnarray}
where 
$$
F_1(1) = 2 \zeta_3 \zeta_2 - \zeta_5 \sim 2.9176809 \cdots \;.
$$
In this way, at the order of {\bf weight 5}, one new 
function\footnote{Compare with results of \cite{HypExp}.},
$F_1$, 
which is not expressible in terms of Nielsen polylogarithms, 
is generated by the Laurent expansion of a Gauss hypergeometric 
function with integer values of parameters. 
In general, the explicit form of this function is not 
uniquely determined, 
and the result may be presented in another form by using a
different subset of harmonic polylogarithms.

\subsubsection{Half-integer values of of $\ep$-independent parameters}
\label{half}

Let us apply a similar analysis for the second basis hypergeometric function
\begin{eqnarray}
&& 
{}_{2}F_1\left(\begin{array}{c|}
a_1 \ep, a_2\ep\\
\frac{1}{2} \!+\! f \ep  \end{array} ~z \right) \; .
\label{gauss:1a}
\end{eqnarray}
In this case, the differential equation has the form
\begin{eqnarray}
\frac{d}{dz} 
\left( z \frac{d}{dz} -\frac{1}{2} + f \ep   \right) w(z)
= 
\left( z \frac{d}{dz} + a_1 \ep \right) \left( z \frac{d}{dz} + a_2 \ep\right) w(z) \;,
\label{gauss:diff:a}
\end{eqnarray}
with the same boundary conditions 
$w(0)=1$ and $\left. z \frac{d}{dz} w(z)\right|_{z=0} = 0$.
Using the $\ep$-expanded form of the solution, and noting that Eq.~(\ref{epsilon-expansion}), 
and in fact, Eq.~(\ref{gauss:diff:a}) is valid at each order of the $\ep$-expansion, 
we may rewrite Eq.~(\ref{gauss:diff:a}) as 
\begin{eqnarray}
\left[ (1-z) \frac{d}{dz} - \frac{1}{2z} \right] \left( z \frac{d}{dz}\right) w_i(z) 
= 
\left[ (a_1 \!+\! a_2) \!-\! \frac{f}{z} \right] \left( z \frac{d }{d z} \right)  w_{i-1} (z)
\!+\! a_1a_2 w_{i-2}(z) \;.
\label{gauss:diff:2a}
\end{eqnarray}

Let us introduce the new variable $y$ such that \footnote{The form of this variable follows from the analysis performed 
in Refs.\ \cite{DK01,DK04,MKL04}.}, 
\begin{eqnarray}
y = \frac{1-\sqrt{\frac{z}{z-1}}}{1+\sqrt{\frac{z}{z-1}}} \;, 
\quad
z = -\frac{(1-y)^2}{4y} \;, 
\quad
1-z = \frac{(1+y)^2}{4y} \;, 
\quad
z \frac{d}{dz} = - \frac{1-y}{1+y} y \frac{d}{dy}\;,
\label{conformal}
\end{eqnarray}
and define a set of a new functions $\rho_i(y)$ so that\footnote{
We may note that 
$$
{}_{2}F_1\left(\begin{array}{c|}
1+a_1\ep, 1\!+\!a_2\ep\\
\tfrac{3}{2}  \!+\! f \ep  \end{array} ~z \right) 
= 
\frac{1+2f\ep}{2z} \frac{1-y}{1+y}
\sum_{k=0}^\infty 
\left[ \frac{\rho_{k+2}(y)}{a_1 a_2} \right]
\ep^k \;.
$$
} 
\begin{equation}
z \frac{d}{dz} w_i(z) \equiv
\left( 
-\frac{1-y}{1+y} y \frac{d}{dy}
\right) w_i(y)  = 
\frac{1-y}{1+y} \rho_i(y) \;,
\label{rho:a}
\end{equation}
and, as in the previous case, 
\begin{equation}
\rho(y) = z \frac{d}{dz} w(z) = \sum_{k=0}^\infty \rho_k(y) \ep^k \;.
\end{equation}
In terms of the new variable $y$, Eq.~(\ref{gauss:diff:2a}) can be 
written as system of two first order differential equations:
\begin{eqnarray}
y \frac{d}{dy} \rho_i (y)
& = &  
\left(a_1 \!+\! a_2 \right) \frac{1-y}{1+y} \rho_{i-1}(y)
+ 2f \left( 
\frac{1}{1-y}
- 
\frac{1}{1+y}
\right) \rho_{i-1}(y) 
\!+\! a_1 a_2 w_{i-2}(y) \;,
\nonumber \\ 
y \frac{d}{dy} w_k(y) & = & - \rho_k(y) \;.
\label{gauss:diff3a}
\end{eqnarray}
The solution of these differential equations for 
functions $w_i(y)$ and $\rho_i(y)$ has the form
\begin{eqnarray}
\rho_i(y) 
& = & 
\int_1^y dt 
\left[ 
2f \frac{1}{1-t} 
\!-\! 2 (a_1\!+\!a_2\!-\!f) \frac{1}{1+t}\right] \rho_{i-1}(t)
- (a_1 \!+\! a_2) \left[ w_{i-1}(y) \!-\! w_{i-1}(1) \right]
\nonumber \\ && 
+ a_1 a_2 \int_1^y \frac{dt}{t} w_{i-2}(t) \;, \quad i \geq 1 \;,
\nonumber \\ 
w_i(y) & = & - \int_1^y \frac{dt}{t} \rho_i(t) \;, \quad i \geq 1 \;.
\label{w:a}
\end{eqnarray}
The point $z=0$ transforms to 
the point $y = 1$ under the transformation (\ref{conformal}),
so that the boundary conditions are 
\begin{eqnarray}
\begin{array}{cl}
 w_k(1) = 0\;,  & k \geq 1 \;, \\
\rho_k(1) = 0\;,  & k \geq 0 \;.
\label{boundary:a}
\end{array} 
\end{eqnarray}

The first several coefficients of the $\ep$-expansion 
can be calculated quite easily by using $w_0(y)=1$ and $\rho_0(y)=0$:
\begin{subequations}
\label{first:a}
\begin{eqnarray}
\rho_1(y) &=& w_1(y) = 0, \\
\frac{\rho_2(y)}{a_1 a_2} &=& \ln(y) \equiv H(0;y) \;,\\
\frac{w_2(y)}{a_1 a_2} &=& -\frac{1}{2} \ln^2 (y) \equiv - H(0,0;y) \;.
\end{eqnarray}
\end{subequations}
Continuing these iterations, we may reproduce the coefficients of the
$\ep$-expansion of the Gauss hypergeometric 
function (\ref{gauss:1a}) presented in  Eq.~(4.2) of Ref.\ \cite{MKL06}.
Since the length of the expressions obtained for the coefficient functions 
$
\rho_3(y), \omega_3(y), 
\rho_4(y), \omega_4(y), 
\rho_5(y), \omega_5(y)
$
is similar to those published in Eq.~(4.1) of Ref.\ \cite{MKL06}, 
we don't reproduce them here.\footnote{M.Y.K. thanks to M.~Rogal for 
pointing out a mistake in Eq.~(4.1) of Ref.\ \cite{MKL06}: In 
the $\ep^2$ term, the coefficient should be ``$-2(3 f-a_1-a_2)$'' 
instead of ``$-2(3f-2a_1-2a_2)$''.} 
%
The higher order terms of $\ep$-expansion are relatively lengthy and 
therefore will also not be presented here. 
Unfortunately, as in the previous case, we are unable to calculate the 
$k$-coefficient of $\ep$-expansion 
without knowledge of previous ones. 

From representation (\ref{w:a}) we deduce the following result:
\begin{itemize}
\item
{\bf Corollary 4:}
{\it 
The all-order $\ep$-expansion of function (\ref{gauss:1a}) can be written in terms of 
harmonic polylogarithms $H_{\vec{A}}(y)$ of variable $y$ defined in (\ref{conformal})
and multiple index $\vec{A}$ with entries taking values $0$, $1$ and $-1$. 
}
\end{itemize}
This statement follows from the representation (\ref{w:a}), 
the values of coefficients functions $w_k(z)$, $k=0,1,2$ (see Eqs.~(\ref{boundary}), (\ref{first})), 
properties of harmonic polylogarithms, and the relation between powers of logarithms and harmonic polylogarithms.
Also, {\bf Corollary 2} and {\bf Corollary 3} are valid for the hypergeometric function (\ref{gauss:1a}).

We would like to mention that, in contrast to the Eq.~(\ref{w}), 
Eq.~(\ref{w:a}) contains a new type of function, coming from the integral $\int f(t) dt/(1+t)$.
Another difference is that the first nontrivial coefficient function, $\rho_2(y)$,
is equal to $\ln(y)$, instead of $\ln(1-z)$, as it was in the previous case. 
It was shown in Ref.\ \cite{RV00} that terms containing the logarithmic singularities can be explicitly factorised
(see Eqs.~(21)-(22) in Ref.\ \cite{RV00}), so that the coefficient functions, 
$w_k(y)$ and $\rho_k(y)$ from Eq.~(\ref{w:a}),  have the form
\begin{eqnarray}
w_k(y) & = & \sum_{j=0}^k 
c(\vec{s}, \vec{\sigma},k)
\ln^{k-j}(y) 
\left[
\Li{\left( \vec{\sigma} \atop \vec{s} \right)}{y}   
- 
\Li{\left( \vec{\sigma} \atop \vec{s} \right)}{1} \right] \;, 
\nonumber \\ 
\rho_k(y) & = & \sum_{j=0}^{k-1}
\tilde{c}(\vec{s}, \vec{\sigma},k)
\ln^{k-j}(y) 
\left[
\Li{\left( \vec{\sigma} \atop \vec{s} \right)}{y}   
- 
\Li{\left( \vec{\sigma} \atop \vec{s} \right)}{1} \right] \;, 
\end{eqnarray}
where $c(\vec{s}, \vec{\sigma},k)$ and $\tilde{c}(\vec{s}, \vec{\sigma},k)$ are numerical coefficients, 
$\vec{s}$ and $\vec{\sigma}$ are multi-index, 
$\vec{s}=(s_1, \cdots s_n)$ and $\vec{\sigma} = (\sigma_1, \cdots, \sigma_n)$, 
$\sigma_k$ belongs to the set of the square roots of unity,
$\sigma_k = \pm 1$,
and $\Li{\left( \vec{\sigma} \atop \vec{s} \right)}{y}$ is 
a coloured multiple polylogarithm of one variable \cite{Goncharov,Broadhurst:1998,Borwein:1999}, 
defined as
\begin{equation}
\Li{\left( \sigma_1, \sigma_2, \cdots, \sigma_k \atop s_1, s_2, \cdots, s_n \right)}{z} 
= 
\sum_{m_1 > m_2 > \cdots m_n > 0} z^{m_1} \frac{\sigma_1^{m_1} \cdots \sigma_n^{m_n}
                                            }{m_1^{s_1} m_2^{s_2} \cdots m_n^{s_n}} \;.
\label{colored}
\end{equation}
It has an iterated integral representation w.r.t.\ three differential forms, 
\begin{eqnarray}
\omega_0 & = & \frac{dy}{y}, \quad \sigma=0,
\nonumber \\ 
\omega_\sigma & = & \frac{\sigma dy}{1- \sigma y}, \quad \sigma= \pm 1,
\end{eqnarray}
so that, 
\begin{equation}
\Li{\left( \sigma_1, \sigma_2, \cdots, \sigma_k \atop s_1, s_2, \cdots, s_k \right)}{y} 
= \int_0^1 
\omega_0^{s_1-1} \omega_{\sigma_1}
\omega_0^{s_2-1} \omega_{\sigma_1 \sigma_2}
\cdots
\omega_0^{s_k-1} \omega_{\sigma_1 \sigma_2 \cdots \sigma_k}
\;,  \quad \sigma_k^2 = 1\;.
\label{color}
\end{equation}
The values of coloured polylogarithms of unit argument were studied in Refs.\ \cite{Borwein1996,color}.

\subsection{Zero-values of the $\ep$-dependent part of upper parameters}
In the case when one of the upper parameter of the Gauss hypergeometric function
is a positive integer, the result of the reduction has the simpler form 
(compare with Eq.~(\ref{decomposition})): 
\begin{eqnarray}
&& \hspace{-5mm}
P(b,c,z)
{}_{2}F_{1}(I_1,b+I_2;c+I_3;z)
= 
Q_1(b,c,z) {}_{2}F_{1}(1,b;c; z) + Q_2(b,c,z) \;,
\label{decomposition:integer}
\end{eqnarray}
where 
$b,c,$ are any fixed numbers,  
$P,Q_1,Q_2$ are polynomial in parameters $b,c$ and argument $z$, 
and $I_1,I_2,I_3$ are any integers.\footnote{The proper 
algebraic relations for the reduction are given in Ref.\ \cite{MKL06}.} 
In this case, it is enough to consider the following two basis functions:
${}_2F_{1}(1, 1+a\ep; 2+c \ep;z)$ and 
${}_2F_{1}(1, 1+a\ep; \frac{3}{2}+f \ep;z)$.
The $\ep$-expansion of this function can be derived from the proper solution 
given by Eq.~(\ref{first}) or Eq.~(\ref{first:a}), using the relations
\begin{eqnarray}
\vspace{-5mm}
{}_{2}F_1\left(\begin{array}{c|}
1, 1\!+\!a_2\ep\\
2  \!+\! f \ep  \end{array} ~z \right) 
& = & 
\lim_{a_1 \to 0 }
\frac{1+c\ep}{a_1 a_2 \ep^2}
\frac{d}{dz}
{}_{2}F_1\left(\begin{array}{c|}
a_1 \ep, a_2\ep\\
1 \!+\! c \ep  \end{array} ~z \right) 
= 
\frac{1+c\ep}{z}
\sum_{k=0}^\infty 
\left[ 
\left. \frac{\rho_{k+2}(z)}{a_1 a_2} \right|_{a_1 = 0}
\right]
\ep^k 
\nonumber \\
\label{integer_1}
\end{eqnarray}
and 
\begin{eqnarray}
\hspace{-3mm}
{}_{2}F_1\left(\begin{array}{c|}
1, 1\!+\!a_2\ep\\
\frac{3}{2} \!+\! f \ep  \end{array} ~z \right) 
& = & 
\lim_{a_1 \to 0 }
\frac{1 \!+\! 2f\ep}{2 a_1 a_2 \ep^2}
\frac{d}{dz}
{}_{2}F_1\left(\begin{array}{c|}
a_1 \ep, a_2\ep\\
\frac{1}{2} \!+\! f \ep  \end{array} ~z \right) 
= 
\frac{1+2f\ep}{2z}
\sum_{k=0}^\infty 
\left[ 
\left. \frac{\rho_{k+2}(y)}{a_1 a_2} \right|_{a_1 = 0}
\right]
\ep^k \;,
\nonumber \\
\label{integer_1a}
\end{eqnarray}
where we have used the differential relation
$$
\frac{d}{dz}
\;{}_{2}F_1\left(\begin{array}{c|}
a, b\\
c \end{array} ~z \right) 
= \frac{ab}{c} 
\;{}_{2}F_1\left(\begin{array}{c|}
1+a, 1+b\\
1+c \end{array} ~z \right) \;,
$$
and 
the brackets mean that in the proper solution, we can put $a_1=0$.
The functions $\rho_k$ are  given by Eq.~(\ref{first}) and Eq.~(\ref{first:a}), correspondingly.
Due to {\bf Corollary  3}, the limit $a_1 \to 0$ must exist. 

The case when both upper parameters are integers may be handled in a 
similar manner. 

{{\bf Theorem 1} is thus proved.} $\blacksquare$

\section{Some particular cases}
\subsection{The generalized log-sine functions and their generalization}
\label{imaginary}
For the case $0 \leq z \leq 1$ 
the variable $y$ defined in (\ref{conformal}) belongs to a complex unit circle, $y=\exp (i \theta)$. 
In this case, the harmonic polylogarithms can be split into 
real and imaginary parts (see the discussion in Appendix A of Ref.\ \cite{MKL04}), 
as in the case of classical polylogarithms. \cite{Lewin}
Let us introduce the trigonometric parametrization
$z = \sin^2 \tfrac{\theta}{2}.$
In this case, the solution of the proper differential equations 
(\ref{w}) and (\ref{w:a})
can be written in the form 
\begin{eqnarray}
\rho_i(\theta) 
& = & 
(a_1 \!+\! a_2 \!-\! c) 
\int_0^\theta d \phi
\frac{\sin \frac{\phi}{2}}{\cos \frac{\phi}{2}} \rho_{i-1}(\phi)
\!+\! a_1 a_2 \int_0^\theta d \phi \frac{\sin \frac{\phi}{2}}{\cos\frac{\phi}{2}} w_{i-2}(\phi) 
\!-\! c w_{i-1}(\theta) \;, \quad i \geq 1 \;,
\nonumber \\ 
w_i(\theta) & = & \int_0^\theta d \phi 
\frac{\cos \frac{\phi}{2}}{\sin \frac{\phi}{2}} \rho_i(\phi) \;, \quad i \geq 1 \;,
\label{w:geom}
\end{eqnarray}
and 
\begin{eqnarray}
\rho_i(\theta) 
& = & 
(a_1 \!+\! a_2 \!-\! f) 
\int_0^\theta d \phi
\frac{\sin \frac{\phi}{2}}{\cos \frac{\phi}{2}} \rho_{i-1}(\phi)
\!-\! f \int_0^\theta d \phi \frac{\cos \frac{\phi}{2}}{\sin \frac{\phi}{2}}  \rho_{i-1}(\phi)
\!+\! a_1 a_2 \int_0^\theta d \phi w_{i-2}(\phi) \;, 
\nonumber \\ 
w_i(\theta) & = & \int_0^\theta d \phi \rho_i(\phi) \;, \quad i \geq 1 \;,
\label{w:a:geom}
\end{eqnarray}
respectively.
In the first case, the solutions of the system of equations (\ref{w:geom})  are 
harmonic polylogarithms with argument equal to $\sin^2 \tfrac{\theta}{2}$.
In the second case, the result contains
the generalized log-sine functions \cite{Lewin,FK99,D00,lsjk} 
and some of their generalizations studied in Ref.\ \cite{MKL05} (see also Ref.\ \cite{Anastasiou}).
For illustration, we will present a first 
several terms of the $\ep$-expansion~\footnote{The FORM representation of these expressions can be extracted from \cite{MKL}.} 
(see the proper relations, Table I of Appendix C in Ref.\ \cite{DK04}):
\begin{eqnarray}
&& 
\left. _2F_1\left( \begin{array}{c} 1+a_1 \ep, 1 + a_2 \ep \\
                 \frac{3}{2} + f \ep \end{array} \right| \sin^2 \tfrac{\theta}{2} \right) 
= 
\frac{(1+2f\ep)}{ \sin \theta} 
\nonumber \\ && \hspace{1mm}
\times
\Biggl(
\theta
+ 2 \ep \Biggl\{
  \gamma_f \left( \Ls{2}{\pi \!-\! \theta} \!-\! \theta L_{\theta} \right) 
- f  \left( \Ls{2}{\theta} \!+\! \theta l_{\theta} \right) 
\Biggr\}
\nonumber \\ && \hspace{1mm}
+ \ep^2 \Biggl\{
  2 f \gamma_{2f} \Ls{3}{\theta}
\!+\! 2 \gamma_f \gamma_{2f} \Ls{3}{\pi-\theta}
\!-\! f \gamma_f \Ls{3}{2 \theta}
\nonumber \\ && \hspace{7mm}
\!+\! 4 f \gamma_f \left[
  \Ls{2}{\theta} L_\theta
\!-\! \Ls{2}{\pi \!-\! \theta} l_\theta
\!+\! \theta L_\theta l_\theta
\right]
+ 4 f^2 \Ls{2}{\theta} l_\theta
- 4 \gamma_f^2 \Ls{2}{\pi-\theta} L_\theta 
\nonumber \\ && \hspace{7mm}
+ 2 f^2 \theta l^2_\theta
+ 2 \gamma_f^2 \theta L_\theta^2 
+ \frac{1}{6} a_1 a_2 \theta^3
+ \gamma_f \gamma_{2f} \pi \zeta_2
\Biggr\}
\nonumber \\ && \hspace{1mm}
+ \ep^3 \Biggl\{
\frac{4}{3} \gamma_{2f}
\left[ 
 (a_1+a_2) \gamma_f  \Ls{4}{\pi-\theta}
+ f^2 \Ls{4}{\theta}
- 3 f \gamma_f \Lsc{2,3}{\theta}
\right]
- \frac{2}{3} f^2 \gamma_{f} \Ls{4}{2\theta}
\nonumber \\ && \hspace{7mm}
+ 4 a_1 a_2 \left[ 
  2 f \Cl{4}{\theta} 
- 2  \gamma_f \Cl{4}{\pi-\theta} 
- f \Cl{3}{\theta} \theta
- \gamma_f \Cl{3}{\pi-\theta} \theta
\right]
\nonumber \\ && \hspace{7mm}
+ 2 \left[ f l_\theta + \gamma_f L_\theta \right]
\left[
  f \gamma_f \Ls{3}{2\theta}
- 2 \gamma_f \gamma_{2f} \Ls{3}{\pi-\theta} 
- 2 f \gamma_{2f} \Ls{3}{\theta} 
\right]
\nonumber \\ && \hspace{7mm}
+ 2 \left[ f l_\theta + \gamma_f L_\theta \right]^2
\left[ 
  2\gamma_{2f} \Ls{2}{\pi-\theta} 
- f \Ls{2}{2\theta} 
\right]
+ a_1 a_2 \gamma_{2f} \theta^2 \Ls{2}{\pi-\theta} 
\nonumber \\ && \hspace{7mm}
- \frac{1}{2} a_1 a_2 f \Ls{2}{2\theta} \theta^2
\!-\! \frac{1}{3} a_1 a_2 \theta^3 \left[ f l_\theta \!+\! \gamma_f L_\theta \right]
\!-\! \frac{4}{3} \theta \left[ f l_\theta \!+\! \gamma_f L_\theta \right]^3
\!-\! 2 \gamma_f \gamma_{2f} \pi \zeta_2 \left[ f l_\theta \!+\! \gamma_f L_\theta \right]
\nonumber \\ && \hspace{7mm}
+  a_1 a_2 (3 a_1 + 3 a_2 - 7 f) \theta \zeta_3
- 2 (a_1 + a_2 )\gamma_f \gamma_{2f} \pi \zeta_3
\Biggr\}
+ {\cal O} (\ep^4)
\Biggr) 
\label{A_expansion:1}
\end{eqnarray}
and 
\begin{eqnarray}
&& \hspace{-5mm}
\left. _2F_1\left( \begin{array}{c} a_1 \ep, a_2 \ep \\
                 \frac{1}{2} + f \ep \end{array} \right| \sin^2 \tfrac{\theta}{2} \right) 
= 
1 + a_1 a_2 \ep^2 
\Biggl(
\frac{1}{2} \theta^2 
\nonumber \\ && \hspace{1mm}
+ \ep \Biggl\{
  2 f \Ls{2}{\theta} \theta
\!-\! 2 \gamma_f \Ls{2}{\pi-\theta} \theta
\!+\! 4 \gamma_f \Cl{3}{\pi-\theta}
\!+\! 4 f \Cl{3}{\theta}
\!+\! (3a_1 \!+\! 3a_2 \!-\! 7f) \zeta_3
\Biggr\}
\nonumber \\ && \hspace{1mm}
+ \ep^2 \Biggl\{
  2 \gamma_f \gamma_{2f} \Ls{3}{\pi-\theta} \theta 
\!-\! f \gamma_f \Ls{3}{2\theta} \theta 
\!+\! 2 f \gamma_{2f} \Ls{3}{\theta} \theta
\!+\! \frac{1}{24} a_1 a_2 \theta^4
\nonumber \\ && \hspace{7mm}
- 2 \left[ f \Ls{2}{\theta} \!-\! \gamma_{f} \Ls{2}{\pi-\theta} \right]^2
+ \gamma_f \gamma_{2f} \theta \pi \zeta_2 
\Biggr\}
+ {\cal O} (\ep^3)
\Biggr) \;, 
\label{A_expansion:2}
\end{eqnarray}
where 
$$
L_\theta = \ln\left( 2 \cos \frac{\theta}{2} \right) \;, 
\quad 
l_\theta = \ln\left( 2 \sin \frac{\theta}{2} \right) \;, 
$$
the generalized log-sine function is defined as 
\begin{equation}
\LS{j}{k}{\theta} =   - \int\limits_0^\theta {\rm d}\phi \;
   \phi^k \ln^{j-k-1} \left| 2\sin\frac{\phi}{2}\right| \, ,
\quad 
\Ls{j}{\theta} = \LS{j}{0}{\theta} \; ,
\label{log-sine}
\end{equation}
and we use the notation $\Lsc{2,3}{\theta}$ for the special combination (see Eq.~(2.18) in Ref.\ \cite{DK04})
\begin{eqnarray}
\label{Lsc-Ti}
\Lsc{2,3}{\theta} &=& \tfrac{1}{12}\Ls{4}{2\theta}
- \tfrac{1}{3}\Ls{4}{\theta}
+ 2 \Ti{4}{\tan\tfrac{\theta}{2}}
- 2 \ln\left( \tan\tfrac{\theta}{2} \right)\;
 \Ti{3}{\tan\tfrac{\theta}{2}}
\nonumber \\ &&
+ \ln^2\left( \tan\tfrac{\theta}{2} \right)\;
 \Ti{2}{\tan\tfrac{\theta}{2}}
- \tfrac{1}{6} \theta \ln^3\left( \tan\tfrac{\theta}{2} \right) \; ,
\end{eqnarray}
where the functions $\Ti{N}{z}$ are defined as \cite{Lewin}
\begin{equation}
\label{Ti_N}
\Ti{N}{z} 
= {\rm Im}\left[ \Li{N}{{\rm i}z}\right]
= \frac{1}{2  {\rm i} } 
\Bigl[\Li{N}{ {\rm i}z} - \Li{N}{- {\rm i}z}\Bigr] \;,
\qquad
\Ti{N}{z} = \int\limits_0^z \frac{{\rm d}x}{x}\; \Ti{N-1}{x} \; .
\end{equation}

These functions receive special interest in physics through their
role in the so-called ``single-scale'' diagrams, 
which depend only on one massive scale parameter. 
The massless propagator-type diagrams, bubble-type diagrams 
and propagator-type diagrams on mass shell all belong to this class.
In particular, the single-scale diagrams with two massive particle 
cuts correspond to 
hypergeometric functions with value of argument equal to $z=1/4$. \cite{FKK,KV00,DK01}
In this case, the value of the conformal variable $y$ is equal to the primitive 
``sixth root of unity'', $y=\exp\left( i \frac{\pi}{3} \right)$.
In contrast to the case in multiple polylogarithms (\ref{mp})  
of the primitive sixth root of unity
studied in Ref.\ \cite{Borwein:2000} and the more complicated case in
coloured polylogarithms of  the
sixth root of unity studied by Broadhurst in Ref.\ \cite{Broadhurst:1998}, 
the physically interesting case corresponds  to coloured polylogarithms of 
square root (\ref{color}) (harmonic polylogarithms) with argument equal to  
primitive sixth root of unity. 
In this case, some new  transcendental constants, in addition to  studied in 
Ref.\ \cite{Borwein:2000} will be generated. 
The set of independent constants up to {\bf weight 5} was constructed in Refs.\ \cite{FK99,DK01,MKL05}.

\subsection{Special cases:  all-order $\ep$-expansion in terms of Nielsen polylogarithms}
\label{special}
One advantage of a trigonometric representation used in the previous section is the theorem 
proved in Ref.\ \cite{DK01} (see also Ref.\ \cite{lsjk}), that any generalized log-sine function (\ref{log-sine})
is expressible in term of Nielsen polylogarithms \cite{Nielsen} only. 
Using this theorem, it was shown in Ref.\ \cite{DK01,DK04} that for the Gauss hypergeometric function 
\begin{equation}
_2F_1 \left(\begin{array}{c|} 
1, 1 \!+\! a\ep  \\
\tfrac{3}{2} \!+\! b\ep\end{array} ~\sin^2 \tfrac{\theta}{2} \right) \;, 
\end{equation}
the Laurent expansion is expressible in terms of only Nielsen polylogarithms
in the three cases (i) $b=0$, (ii) $b=a$, (iii) $a=2b$.
Using the reduction algorithm \cite{MKL06}, we can claim that the
Laurent expansions of the following functions are also expressible in terms 
of Nielsen polylogarithms only:
\begin{eqnarray}
_2F_1 \left(\begin{array}{c|} 
I_1, I_2 \!+\! \ep  \\
\tfrac{1}{2} \!+\! I_3 \end{array} ~\sin^2 \tfrac{\theta}{2} \right)\; , 
\quad 
_2F_1 \left(\begin{array}{c|} 
I_1, I_2 \!+\! \ep  \\
\tfrac{1}{2} \!+\! I_3 \!+\! \ep \end{array} ~\sin^2 \tfrac{\theta}{2} \right) \;, 
\quad
_2F_1 \left(\begin{array}{c|} 
I_1, I_2 \!+\! \ep  \\
\tfrac{1}{2} \!+\! I_3 \!+\! \tfrac{1}{2}\ep \end{array} ~\sin^2 \tfrac{\theta}{2} \right) \;, 
\label{caseI}
\end{eqnarray}
where $I_1$, $I_2$ and $I_3$ are integers. 

It is interesting to analyze this solution from the point of view of 
Eq.~(\ref{w:a:geom}). 
Due to fact that $a_1 = 0$, the last term in Eq.~(\ref{w:a:geom}) is
identically equal to zero. In case (i), only the first term survives, 
with integration kernel having the form $d \ln(\cos \tfrac{\phi}{2})$. 
In case (ii), only the second term survives, and the
integration kernel has the form $d \ln(\sin \tfrac{\phi}{2})$. 
In case (iii), the first and second terms can be reduced to the 
second case of a double argument.  The 
statement about expressibility of {\it inverse binomial sums} in terms 
of log-sine function, proved in Ref.\ \cite{DK04} (see also Ref.\ \cite{MKL04})
 applies to all three of these cases.

We can extend the class of Gauss functions whose $\ep$-expansions
are expressible in terms of only Nielsen polylogarithms by using 
algebraic relations\footnote{This can also be derived via the integral representation.}
between of the fractional-linear arguments 
(see Sec.\ 3 in Ref.\ \cite{MKL06}). The cases which may be expressed in this
manner are summarized in {Table I}, where $a,b,c$ are parameters 
of the Gauss hypergeometric functions ${}_2F_1(a,b;c;z)$ and 
$I_1$, $I_2$ and $I_3$ are integer:
$$
\begin{array}{|c|c|c|c|c|c|c|c|c|c|}\hline
\multicolumn{10}{|c|}{\rm\bf{Table\ \ I}    } \\   \hline
a & I_1                                & I_1 & I_1 & I_1 & I_1 & I_1 
  & I_1 \!+\! \ep & I_1 \!+\! \ep & I_1 \!+\! 2 \ep \\ \hline
b & \frac{1}{2} \!+\! I_2              & \frac{1}{2} \!+\! I_2 \!+\! \ep  & \frac{1}{2} \!+\! I_2 \!-\! \ep 
  & \frac{1}{2} \!+\! I_2 \!+\! \ep    & \frac{1}{2} \!+\! I_2            & \frac{1}{2} \!+\! I_2 \!+\! \ep 
  & \frac{1}{2} \!+\! I_2   & \frac{1}{2} \!+\! I_2 \!+\! \ep  & \frac{1}{2} \!+\! I_2 \!+\! \ep \\ \hline
c & \frac{1}{2} \!+\! I_3 \!+\! \ep    & \frac{1}{2} \!+\! I_3            & \frac{1}{2} \!+\! I_3 \!+\! \ep
  & I_3 \!+\! \ep                      & I_3 \!+\! \ep                    & I_3 \!+\! 2 \ep       
  & I_1 \!+\! 1\!+\! \ep & I_1 \!+\! 1 \!+\! \ep & I_1 \!+\! 1 \!+\! 2 \ep \\ \hline
\end{array}
$$

The results of this section can be formulated as follows: 

\begin{itemize}
\item
{\bf Proposition 1}: {\it
All cases of Gauss hypergeometric functions with half-integer values of 
parameters for which the all-order
$\ep$-expansion is expressible in terms of only Nielsen polylogarithms 
are described in Eq.\ (\ref{caseI}) or the parameters shown in Table I. }
\end{itemize}


\section{Conclusions}
The main result of this paper is the proof of {\bf Theorem 1}, 
as stated also in the abstract.  The proof includes two steps:
(i) the algebraic reduction of Gauss hypergeometric functions of the type
in {\bf Theorem 1} to basic
functions and (ii) the iterative algorithms for calculating the analytical
coefficients of the $\ep$-expansion of basic hypergeometric functions. 

In implementing step (i), the algebraic relations between basis 
functions with half-integer values of parameters reduce all of the cases to
the one basic function of type (\ref{2F1-type1}) and its first derivative
(see details in Ref.\ \cite{MKL06}).
In step (ii), the algorithm is constructed for integer values of parameters 
in Eq.~(\ref{w}) and for basis Gauss hypergeometric functions with half-integer 
values of parameters in Eq.~(\ref{w:a}).
This allows us to calculate the coefficients directly, without reference to
 multiple sums.  

It is interesting to note that the Laurent expansions of the
 Gauss hypergeometric functions with integer values of parameters are 
expressible in terms of multiple polylogarithms of one variable
(see Eq.~(\ref{mp})) or the Remiddi-Vermaseren harmonic polylogarithms 
with multiple index including only values $0$ and $1$. 
The argument of the resulting functions coincides 
with the original variable of the hypergeometric function. 
For Gauss hypergeometric functions with half-integer values of parameters, 
the coefficients of the
$\ep$-expansion  produce the full set of harmonic polylogarithms, or 
coloured multiple polylogarithms of one variable (see Eq.~(\ref{colored})). 
These functions depend on a new variable, related to
the original variable by conformal transformation (see Ref.\  \cite{MKL06}).

For special values of the argument of the hypergeometric function, $z<1$, 
the coloured multiple polylogarithms of one variable may be split 
into real and imaginary parts. 
This case has been discussed in section \ref{imaginary}.
It was shown that the physically interesting case, representing 
single-scale diagrams with with two massive particle cuts, 
corresponds to coloured polylogarithms (\ref{color}) with argument equal to a primitive 
``sixth root of unity'', $y=\exp\left( i \frac{\pi}{3} \right)$. 
This gives an explanation of the proper 
``basis of transcendental constants'' constructed in Refs.\ \cite{FK99} 
and \cite{DK01}, 
and its difference from the proper basis of David Broadhurst \cite{Broadhurst:1998}.

In the section \ref{special}, the 
subset of Gauss hypergeometric function is analyzed, showing 
that the all-order $\ep$-expansion is expressible in terms of 
Nielsen polylogarithms only. In particular, we have formulated 
the proposition that the
only Gauss hypergeometric functions with half-integer values of parameters
for which the all-order $\ep$-expansion is expressible in terms of 
Nielsen polylogarithms only
belong to one of the functions described in (\ref{caseI}) or in Table I. 

In Appendix \ref{appendix}, we discuss the construction of 
the all-order Laurent expansion of the Gauss 
hypergeometric function (\ref{gauss:1a}) around $z=1$.
\acknowledgments 
We are grateful A.~Davydychev for useful discussion.
M.Yu.K. is thankful to participants of conference ``Motives and Periods'',
University of British Columbia, Vancouver, June 5-12, 2006 \cite{vancouver},
for interesting discussions. Special thanks to Andreas Rosenschon 
for invitation and financial support
and D.~Kreimer and H.~Gangl for enormously and useful discussions and suggestions. 
M.Yu.K. is very grateful to Laura Dolchini for moral support when paper was written.
This research was supported in part by RFBR grant \# 04-02-17192 , 
NATO Grant PST.CLG.980342 and DOE grant DE-FG02-05ER41399.

\appendix 
\section*{Appendix}
\section{The Laurent expansion of Gauss hypergeometric functions
   with half-integer values of parameters around $z=1$}
\label{appendix}

The identities between harmonic polylogarithms (\ref{mp}) under the action of 
the group of fractional-linear transformation of the argument, 
$$
z \to 1-z, \quad \frac{1}{z}, \quad z^2, 
$$
was considered in Ref.\ \cite{RV00} (see also Ref.\ \cite{mp:relation}).
It was shown \cite{RV00} that the full set of Remiddi-Vermaseren functions 
is invariant with respect to transformations 
$$
z \to \frac{1}{z}, \quad \frac{1-z}{1+z}.
$$
In this Appendix, we wish to show that the coefficient functions 
$\omega_k(y)$ and $\rho_k(y)$ entering in the
Laurent expansion of the hypergeometric function (\ref{gauss:1a})
satisfy to some identities with respect to the argument transformation
$z \to 1-z$. 
The derivation is trivial if we recall that the all-order Laurent expansion of  
the hypergeometric function (\ref{gauss:1a}) can be written in terms of 
coloured polylogarithms of argument $y$. 
Under the transformation $z \to 1-z$, the conformal variable simple changes its sign, $y=\to -y$. 
The full set of Remiddi-Vermaseren functions is invariant (up to the addition of a constant imaginary part) 
with respect to changing of the sign of this variable. Consequently, the coefficient functions
$\omega_k(y)$ and $\rho_k(y)$ should be related to
$\omega_k(-y)$ and $\rho_k(-y)$.
%
%

Let us present the explicit relations. 
Using the Kummer's relations between hypergeometric functions
of arguments $z$ and $1-z$, we obtain
\begin{eqnarray}
&& 
{}_{2}F_1\left(\begin{array}{c|}
a_1 \ep, a_2\ep\\
\frac{1}{2} \!+\! f \ep  \end{array} ~1-z \right) 
\nonumber \\ && 
= 
- \frac{z^{1/2+(f-a_1-a_2)\ep}}{ (1-z)^{-1/2+f\ep}}
\frac{\Gamma\left(\frac{1}{2} \!+\! f \ep   \right) \Gamma\left(\frac{1}{2} \!+\! (a_1+a_2-f) \ep\right)}
     {\Gamma\left(1 \!+\! a_1 \ep \right)  \Gamma\left(1 \!+\! a_2 \ep \right)}
\frac{d}{dz}
\;{}_{2}F_1\left(\begin{array}{c|}
-a_1 \ep, -a_2\ep\\
\frac{1}{2} \!+\! (f\!-\!a_1\!-\!a_2) \ep  \end{array} ~z \right) 
\nonumber \\ && \hspace{5mm}
+ \frac{\Gamma\left(\frac{1}{2} \!+\! f \ep   \right) \Gamma\left( \frac{1}{2} \!+\! (f-a_1-a_2) \ep\right)}
     {\Gamma\left(\frac{1}{2} \!+\! (f-a_1) \ep \right)  \Gamma\left(\frac{1}{2} \!+\! (f-a_2) \ep \right)}
\;{}_{2}F_1\left(\begin{array}{c|}
a_1 \ep, a_2\ep\\
\frac{1}{2} \!+\! (a_1+a_2-f) \ep  \end{array} ~z \right) \;.
\label{symmetries}
\end{eqnarray}
The all-order $\ep$-expansion for the hypergeometric functions entering in r.h.s.\ of this relation
is constructed in Sec.~\ref{half}.  Let us apply the same technique 
for constructing the 
Laurent expansion of the hypergeometric function on the l.h.s. 
In accordance with standard procedure (see Ref.\ \cite{Slater}), 
let us introduce a new variable, $Z=1-z$, so that the the differential equation 
around $Z=0$ has the form
\begin{eqnarray}
\frac{d}{dZ} 
\left( Z \frac{d}{dZ} -\frac{1}{2} + (a_1+a_2-f) \ep   \right) w(Z)
= 
\left( Z \frac{d}{dZ} + a_1 \ep \right) \left( Z \frac{d}{dZ} + a_2 \ep\right) w(Z) \;,
\label{gauss:diff:z=1}
\end{eqnarray}
This equation is equivalent to Eq.~(\ref{gauss:diff:a}) with the proper change of variable and one of the parameters,
$$
(z,f) \longleftrightarrow (Z,a_1+a_2-f) \;,
$$
so that we can use the results of Sec.~\ref{half} with the proper change of notations. 
In particular, the solutions of the differential equations for the 
functions $\rho_i(Z)$ and $w_i(Z)$ have the form
\begin{eqnarray}
\rho_i(Y) 
& = & 
\int_1^Y dt 
\left[ 
2 (a_1\!+\!a_2\!-\!f) \frac{1}{1-t} 
\!-\! 2 f \frac{1}{1+t}\right] \rho_{i-1}(t)
- (a_1 \!+\! a_2) \left[ w_{i-1}(Y) \!-\! w_{i-1}(1) \right]
\nonumber \\ && 
+ a_1 a_2 \int_1^Y \frac{dt}{t} w_{i-2}(t) \;, \quad i \geq 1 \;,
\nonumber \\ 
w_i(Y) & = & - \int_1^Y \frac{dt}{t} \rho_i(t) \;, \quad i \geq 1 \;.
\label{w:1-z}
\end{eqnarray}
where  new variable $Y$ is defined as 
\begin{eqnarray}
Y = \frac{1-\sqrt{\frac{Z}{Z-1}}}{1+\sqrt{\frac{Z}{Z-1}}} \equiv -y \;, 
\end{eqnarray}
and $y$ is defined by Eq.~(\ref{conformal}).
In this way, both parts of relation (\ref{symmetries}) are expressible in terms 
of coloured polylogarithms depending on the arguments $y$ (r.h.s.) and 
$-y$ (l.h.s.).

We expect that relations following from  Eq.~(\ref{symmetries})
may be useful in obtaining some dual relations for coloured polylogarithms
(see Ref.\ \cite{Broadhurst:1998,Borwein:1999,Borwein:2000}), 
and in the obtaining algebraic relations between coloured polylogarithms 
of the primitive ``sixth root of unity'', as in case of multiple 
zeta-values \cite{MZV}.
At the present moment, we are not ready to discuss these relations.


\end{document}